\documentclass{article}

 \usepackage[preprint]{neurips_2025}

\usepackage[utf8]{inputenc} % allow utf-8 input
\usepackage[T1]{fontenc}    % use 8-bit T1 fonts
\usepackage{hyperref}       % hyperlinks
\usepackage{url}            % simple URL typesetting
\usepackage{booktabs}       % professional-quality tables
\usepackage{amsfonts}       % blackboard math symbols
\usepackage{nicefrac}       % compact symbols for 1/2, etc.
\usepackage{microtype}      % microtypography
\usepackage{xcolor}         % colors

\usepackage{amsmath}
\usepackage{algorithm}
\usepackage{algpseudocode}
\usepackage{graphicx}
\usepackage{svg}

\title{LLM Agents for Combinatorial Efficient Frontiers: Investment Portfolio Optimization}

\author{%
  Simon Paquette-Greenbaum \quad Jiangbo Yu\thanks{Corresponding author: \texttt{jiangbo.yu@mcgill.ca}}\\
  Department of Civil Engineering, McGill University\\
  Montreal, Quebec, Canada \\
}

\begin{document}

\maketitle

\begin{abstract}
    Investment portfolio optimization is a task conducted in all major financial institutions. The Cardinality Constrained Mean-Variance Portfolio Optimization (CCPO) problem formulation is ubiquitous for portfolio optimization. The challenge of this type of portfolio optimization, a mixed-integer quadratic programming (MIQP) problem, arises from the intractability of solutions from exact solvers, where heuristic algorithms are used to find approximate portfolio solutions. CCPO entails many laborious and complex workflows and also requires extensive effort pertaining to heuristic algorithm development, where the combination of pooled heuristic solutions results in improved efficient frontiers. Hence, common approaches are to develop many heuristic algorithms. Agentic frameworks emerge as a promising candidate for many problems within combinatorial optimization, as they have been shown to be equally efficient with regard to automating large workflows and have been shown to be excellent in terms of algorithm development, sometimes surpassing human-level performance. This study implements a novel agentic framework for the CCPO and explores several concrete architectures. In benchmark problems, the implemented agentic framework matches state-of-the-art algorithms. Furthermore, complex workflows and algorithm development efforts are alleviated, while in the worst case, lower but acceptable error is reported.
\end{abstract}

\section{Introduction}
\subsection{Context}
Agentic Large Language Models (LLM) are emerging as critical elements in automating large workflows and decision support systems in many fields, such as logistics \citep{logistics}, management \citep{management}, healthcare \citep{healthcare}, urban planning \citep{urban}, and transportation \citep{transportation}. Hence, LLM agents have consequently been cast as tools for algorithm development in relevant combinatorial optimization problems such as scheduling \citep{funsearch}. They have been studied extensively in natural language processing for optimization \citep{NL4Opt}, where natural-language problem descriptions have been effectively translated into valuable mathematical formulations. Language model agents are equally proficient at generating algorithmic solutions to optimization problems when presented with natural-language descriptions and mathematical formulations \citep{CoE}, and their performance has also been shown to exceed that of human experts in time-constrained scenarios \citep{COBench}.

Researchers and industry alike have been very interested in these frameworks, as they serve as valuable benchmarks for some of the foremost applications of LLMs. These applications stimulate two crucial facets of LLM performance: Natural-Language Processing (NLP) and coding \citep{agentic}. However, application of language model agents to combinatorial optimization problems has been limited to academic cases, where problem framing and descriptions are taken directly or inspired by textbooks used in human education \citep{ORQA}. Although these studies present structured benchmarks for LLM agents, they are limited to problems with single objectives and are tractable to exact solution approaches.

Limited work has been conducted on combinatorial optimization that reflects real life. Problems that reflect real life are rarely transcribed into textbooks and can seldom be solved exactly, either due to computational resource constraints or problem uncertainty. Furthermore, real-life problems rarely come without trade-offs \citep{MOOP}, and real-life decision makers often require knowledge of those trade-offs' implications, i.e., the Pareto fronts resulting from multi-objective optimization tasks. Studies have applied LLM agents to develop heuristic algorithms for problems intractable to exact solvers \citep{disc_heur}. Still, limited work has been conducted on LLM agent solutions for multi-objective optimization problems.

\subsection{Rationale}
This study's focus is on the development of a language model framework capable of generating algorithm solutions to combinatorial optimization problems, reflecting real-life worst-case scenarios (opposite to several benchmark studies, where problem framing is a best-case scenario). Combinatorial optimization problems are frequently intractable due to their NP-hardness and due to the combinatorial explosion of subsets \citep{comb_np}, rendering exact solutions exploring all subsets intractable. Furthermore, real-life problems are often multi-objective, where decision makers are faced with competing business needs, etc.

An agentic framework capable of heuristic algorithm development presents itself as a valuable tool in such instances.  Problems intractable to exact solvers often require approximate solutions from metaheuristics \citep{exact_heuristic}. Furthermore, multi-objective optimization problems have been shown to sometimes greatly benefit from the pooling of heuristic solutions, where pooled heuristic solutions form non-dominated frontiers with greater convergence and coverage \citep{alg_port}. For example, the Cardinality-Constrained Mean-Variance Portfolio Optimization (CCPO) \citep{CCPO} fits this description very well. Unlike standard Markowitz mean-variance portfolio optimization \citep{MVPO}, which can be solved trivially with exact dynamic programming, CCPO is an NP-hard problem \citep{CCPO_NP} with a non-convex and discontinuous efficient frontier. The subject of metaheuristic CCPO solutions has been studied extensively in the literature \citep{MVPO_review}, and the performance of pooled heuristics in this case has been shown to greatly improve performance beyond singular heuristics \citep{CCPO_hybrid_pool}.

\subsection{Contributions}
Following this rationale, this study contributes several findings:
\begin{itemize}
    \item This study presents an agentic language model framework that serves in the construction of algorithm portfolios for multi-objective combinatorial optimization problems with NP-hardness. The agent framework not only trivializes the immense developmental burden associated with the construction of algorithm portfolios but also has the added benefit of the potential discovery of novel algorithms.
    
    \item This study validates the agentic framework, along with its produced algorithm portfolio, in a series of challenging multi-objective investment portfolio optimization benchmark problems studied extensively throughout the literature. The agent framework, measured against CCPO cases taken from OR-Library \citep{OR-lib}, is shown to produce algorithms on par with the state of the art. Furthermore, the pooling of algorithms from its derived algorithm portfolio is shown to greatly enhance solution performance.
\end{itemize}

\section{Preliminary}
\subsection{LLM Agents}
Coding agents have received a lot of attention within the broader computer science and machine learning communities. They have been shown to be extremely promising in terms of coding ability and simplifying complex workflows. They have also received considerable attention in the domain of combinatorial optimization, given the large algorithm design burden in terms of required effort in solving intractable or NP-hard problems \citep{HeuriGym}. Several coding agent frameworks have been employed successfully in solving combinatorial optimization problems, namely Self-Refine \citep{self-refine}, FunSearch \citep{funsearch}, and ReEvo \citep{ReEvo}.

\subsubsection{Agent Framework}
The coding agent frameworks employed in this study serve as the central element in the development of its algorithm portfolio. Given the successful usage of iterative agents in algorithm generation and discovery, this study will base its framework around their architectures. One-time generation using LLMs has been shown to most often lead to algorithms that frequently exhibit suboptimal performance or runtime/execution errors \citep{self-refine}. Hence, agent frameworks like the one used in this study rely on iterative refinement with external environments, formally referred to as reasoning-action iterations \citep{react}.

The language model, herein referred to as $\mathcal{M}$, will be framed in this study as a coding agent. \autoref{agent} describes the overview of one generation instance of algorithm $\mathcal{A}$ via coding agent $\mathcal{M}$, where $\mathbf{p}$ is the vector of engineered prompt templates, and $a$, $f$, and $s$ are the algorithm(s), feedback(s), and the score(s) from the previous iteration(s).
\begin{align}
    \mathcal{M}(\mathbf{p}, a, f, s) \;\mapsto\; \mathcal{A} \label{agent}
\end{align}

This study's agent framework is based on a greedy refinement agent framework as described in \citep{COBench}, which itself is based on Self-Refine \citep{self-refine}, where greedy refinements are iteratively used to generate algorithms. \autoref{reasoning_action} describes the reasoning-action iterations of this study's agent framework. Here, reasoning requires the vector of engineered prompt templates $\mathbf{p}=\{p,p_\text{PF},p_\text{RA},p_\text{I/O}\}$, including $p$ for general iteration instructions, $p_{\text{PF}}$ for the problem formulation, $p_{\text{RA}}$ for role assignment, and $p_{\text{I/O}}$ for formatting instructions. Furthermore, $a_t^*$, $f_t^*$, and $s_t^*$ are respectively the algorithm, feedback, and scores of the best-scoring previous iteration in ASCII format at an iteration $t$, injected in the prompt templates where appropriate. Action requires problem parameters and inputs $x$ (constant over iterations), as well as the generated algorithm $\mathcal{A}$.
\begin{align}
    \mathcal{A}_{t} \sim \mathcal{M}(\mathbf{p}, a_{t-1}^*, f_{t-1}^*, s_{t-1}^*) \quad \& \quad \{f_{t},s_{t}\}\sim\mathcal{A}_{t}(x) \label{reasoning_action}
\end{align}

\subsubsection{Scoring}
In this study, algorithm solutions are scored externally to the language model agent $\mathcal{M}$, eliminating the possibility of hallucinated and biased self-assessments. Generated algorithm $\mathcal{A}$ is tasked with producing metaheuristic solutions to a multi-objective combinatorial optimization problem. Hence, the action portion of reasoning-action iterations requires the implementation of metrics beyond simple objective fitness scores. Here, the decision makers studying multi-objective optimization problems rely on information pertaining to the efficient frontier of solutions, i.e., the set of non-dominated solutions nearest optimality. For example, in investment portfolio optimization, no rational decision maker would select a portfolio of assets over another with greater risk and for the same return. Inversely, they would neither select a portfolio over another with a lesser return with the same risk. This study defines the efficient frontier, or more formally, the Pareto front, as the subset $H \subseteq Y$ of strictly non-dominated solutions, taken from the set $Y$ of all feasible solutions. \autoref{dominance} defines how a point $y' \in Y$ strictly dominates another point $y \in Y$ across objective dimensions $n$, and \autoref{H} defines the subset $H$ of strictly non-dominated solutions.
\begin{align}
 y' \prec y \;\Longleftrightarrow\; y'_i < y_i \quad \forall i \in \{1,\dots,m\} \label{dominance} \\
 H = \{ y \in Y \mid \forall y' \in Y,\; y' \not\prec y \} \label{H}
\end{align}

Several performance metrics can be used to quantify the performance of the generated solutions contained in the set $H$ \citep{MVPO_review}. Convergence and coverage/diversity are the main metrics through which efficient frontier performance is measured. Convergence represents the closeness of the approximate frontier to that of a theoretical Pareto optimal frontier. Coverage represents the uniformity of distribution of approximate solutions obtained along the efficient frontier. Furthermore, domain-specific metrics can be used to measure Pareto front performance. For example, Percentage-deviation Error (PE) has been used commonly in benchmarking CCPO convergence performance \citep{CCPO}. However, this study relies on hybrid metrics that measure both convergence and coverage. The hypervolume indicator \citep{HV} and Inverted Generation Distance (IGD) \citep{IGD} are commonly used to that effect. In this study, IGD is relied upon as the standard Markowitz portfolio optimization provides an Unconstrained Efficient Frontier (UEF) as a reference for scoring. \autoref{IGD} describes the computation of the IGD metric, where $P$, taken as the UEF in this study, represents the set of theoretical optimal Pareto solutions $y^*$, $H$ is the set of non-dominated approximate Pareto solutions $y$, and $\| y^* - y \|$ is the Euclidean distance between approximate solution $y$ and nearest optimal solution $y^*$.
\begin{align}
\text{IGD}(P, H) = \frac{1}{\text{card}(P)} \sum_{y^* \in P} \min_{y \in H} \| y^* - y \| \label{IGD}
\end{align}

\subsection{Problem Formulation}
This study employs the CCPO problem formulation described in \citep{CCPO} as it has been studied extensively in the literature, where state-of-the-art algorithm performance can be mapped as a reference. The CCPO is a constrained version of the standard Markowitz mean-variance portfolio optimization. Like the standard Markowitz approach, variance is assumed to be an adequate measure of the risk associated with the investment portfolio.

\subsubsection{Objective}
CCPO can be cast as a multi-objective optimization problem, where the objectives are to minimize portfolio risk and maximize portfolio return (equivalently, minimize the negative of return) across potential portfolios composed from a universe of $N$ assets. Like the standard Markowitz approach, \autoref{risk} defines risk and is taken as the variance of portfolio return, where $w_i$ is the proportion held of an asset $i$ ($\forall i \in \{1,\ldots,N\}$) and $\sigma_{ij}$ is the covariance between assets $i$ and $j$ ($\forall i \in \{1,\ldots,N\}$ and $\forall j \in \{1,\ldots,N\}$). Return, described in \autoref{return}, is simply the weighted sum of expected return $\mu_i$ of assets $i$ ($\forall i \in \{1,\ldots,N\}$).
\begin{align}
    \min \quad &\sum_{i=1}^{N} \sum_{j=1}^{N} w_i w_j \sigma_{ij} \label{risk}\\
    \min \quad &-\sum_{i=1}^{N} w_i \mu_i \label{return}
\end{align}

Given the computational complexity of the aforementioned CCPO problem, a single objective formulation, as opposed to a multi-objective formulation, is desirable, as it permits the use of single objective metaheuristics, where multi-objective metaheuristics can add immense computational burden to an already challenging problem. Hence, techniques are used to convert multi-objectives into single objectives. Here, the Weighted Sum (WS) method \citep{WS} for objective linear scalarization is taken in favor of the $\varepsilon$-constraint objective formulation \citep{eps_constraint}, for the same reason, the computational complexity of the aforementioned CCPO problem. Unlike the $\varepsilon$-constraint objective formulation, non-convex regions of the efficient frontier are not covered with the WS objective formulation, where, given its linear nature, only solution points on the convex hull of the objective set can be found. However, decision makers concerned with CCPO efficient frontiers do not typically concern themselves with the exact frontier shape but are satisfied with the trade-off information it provides. In this study, the objective makes use of WS for computational efficiency, where non-convex regions are not considered. Optimization problems requiring complete frontier diversity and coverage should instead opt for $\varepsilon$-constraint objective formulation.

In this instance, there is no known \textit{a priori} preference between risk and return objectives. Here, the interest lies in the efficient frontier of investment portfolio solutions. It is necessary to sweep the linear weights of the WS method to find an efficient frontier, as it is unreasonable to assume that exploratory capacity from metaheuristics alone will cover the efficient frontier without guidance towards specific objective trade-off ratios. In the general case for $n$ objectives, the weight vector $\boldsymbol{\lambda}$ can be found by exploring a unit simplex \citep{simplex} of weights. Exploring trade-off ratios is then simply a matter of sweeping the weight simplex $\Lambda^{n-1}$ defined in \autoref{simplex}.
\begin{align}
    \quad & \Lambda^{n-1} = \left\{ \boldsymbol{\lambda} \in \mathbb{R}^n \;\middle|\; 
\lambda_i \ge 0,\; \sum_{i=1}^{n} \lambda_i = 1 \right\} \label{simplex}
\end{align}

Using the risk and return objectives ($n=2$) defined in \autoref{risk} and \autoref{return}, a single objective can be defined formally, where $\boldsymbol{\lambda}=\{\lambda_1,\lambda_2\}$. Furthermore, the objective can be simplified given the linear equality constraint of the unit simplex, where the weight vector becomes $\boldsymbol{\lambda}=\{\lambda,1-\lambda\}$ as described in \autoref{obj}. Exploring trade-off ratios between risk and return becomes simply sweeping between $\lambda\in[0,1]$, where $\lambda=0$ represents the scenario where return is maximized irrespective of risk. Conversely, $\lambda=1$  represents the scenario where risk is minimized irrespective of return.
\begin{align}
    \min \quad 
        & \lambda \left[\sum_{i=1}^{N} \sum_{j=1}^{N} w_i w_j \sigma_{ij} \right]-(1-\lambda)\left[\sum_{i=1}^{N} w_i \mu_i\right] \label{obj}
\end{align}

\subsubsection{Constraints}
This formulation includes cardinality constraints for the number of assets included in the portfolio, where the number of selected assets is fixed to $K$ as described in \autoref{cardinality}. Accordingly, zero-one decision $z$ variables are introduced for assets as described in \autoref{binary}. Boundary constraints are included for the minimum (buy-in threshold) and maximum proportions of assets held, if they are held, where $\varepsilon$ and $\delta$ are respectively the minimum and maximum proportions as described in \autoref{bounds}. Like standard Markowitz portfolio optimization, budget constraints are also enforced as described in \autoref{budget}. Furthermore, transaction costs and round lot constraints are not considered in order to benefit from the plethora of studies relying on the benchmark problems of CCPO.
\begin{align}
        \quad & \sum_{i=1}^{N} w_i = 1 \label{budget} \\
        & \sum_{i=1}^{N} z_i = K \label{cardinality} \\
        & \varepsilon_i z_i \;\le\; w_i \;\le\; \delta_i z_i, 
            \quad \forall i \in \{1,\ldots,N\} \label{bounds} \\
        & z_i \in \{0,1\}, \quad \forall i \in \{1,\ldots,N\} \label{binary}
\end{align}

\section{Methods}
\subsection{Approach}
In this study, an LLM is framed algorithmically as a coding agent, where the Multi-objective Combinatorial-optimization Agent (\textsc{MoCo--Agent}) algorithm, designed in this study, employs reasoning-action iterations \citep{react} in the generation of Python metaheuristic algorithms. \autoref{Moo--Agent} presents the employed \textsc{MoCo--Agent} algorithm in its generalized form, abstracted for any multi-objective optimization problem. The algorithm employs external functions and templates in the generation of prompts that serve as inputs to an internal language model. The \textsc{MoCo--Agent} defined in this study is designed to operate greedily in its reasoning-action iterations, where the performance, represented by feedback $f_t^*$ and score $s_t^*$, of the best-scoring generated Python algorithm $\mathcal{A}_t^*$, is injected into model prompt inputs at every iteration beyond instantiation. In this instance, a greedy approach to reasoning-action iterations is elected, as its simplicity and low token usage have been shown to outperform more complex agentic frameworks \citep{COBench} when employed in algorithm generation tasks for combinatorial optimization.

An initial algorithm $\mathcal{A}_0$ is generated at the initial iteration $t=0$, where model prompts are free from previous algorithm $a$, feedback $f$, and scoring $s$ information. However, engineered prompts for the CCPO problem formulation $p_{\text{PF}}$, role assignment $p_{\text{RA}}$, and Python algorithm formatting instructions $p_{\text{I/O}}$ are still included. CCPO objective and constraints are written in ASCII format in $p_{\text{PF}}$. The \textsc{MoCo--Agent}'s role as a coding agent is defined in $p_{\text{RA}}$, including the instructions pertaining to the type of metaheuristic algorithm to be developed, as well as instructions pertaining to algorithm hybridization. Formatting instructions are defined in $p_{\text{I/O}}$, where strict constraints are imposed on Python code generation such that generated code adheres to external \textsc{MoCo--Agent} algorithm functionalities, such as calling on libraries for loading asset universes.

\begin{figure}[h]
    \centering
    \begin{algorithm}[H]
    \caption{\textsc{MoCo--Agent}}
    \begin{algorithmic}[1]
    
        \Require input $x$, model $\mathcal{M}$, prompts $\mathbf{p}$, optimal front $P$, iterations $T$, dimensions $n$, step size $\Delta\lambda$
        \vspace{0.5em}
    
         \State $\mathcal{A}_0$ = $\mathcal{M}(\mathbf{p},\cdot)$ \Comment{Instantiate algorithm (\autoref{agent})}
         \vspace{0.5em}
         
        \ForAll {objective vectors $\boldsymbol{\lambda} \in \{\Lambda^{n-1} \text{ discretized by } \Delta\lambda\}$} \Comment{Simplex (\autoref{simplex})}
            \State $Y_{\boldsymbol{\lambda}}=\mathcal{A}_0(x,\boldsymbol{\lambda})$
        \EndFor
        \vspace{0.5em}
    
        \State $Y_0$ $=$ \Call{Aggregate}{$Y_{\boldsymbol{\lambda} \in \Lambda}$}
        \vspace{0.5em}
        
        \If{\Call{Feasible}{$Y_0$}} \Comment{Feasibility (e.g., \autoref{budget}-\ref{binary})}
            \State $H_0 \subseteq Y_0$ \Comment{Non-dominated set (\autoref{dominance}-\ref{H})}
            \State $\{f_0,s_0\}=\text{IGD}(P,H_0)$ \Comment{IGD metric (\autoref{IGD})}
        \Else
            \State $\{f_0,s_0\}=\infty$
        \EndIf
        \vspace{0.5em}
        
        \State $\{a^*_{0},f^*_{0},s^*_{0}\}$ $=$ \Call{Textify}{$\mathcal{A}_0,f_0,s_0$} \Comment{Initial \textit{"best"} iteration}
        \vspace{0.5em}
        
        \For{iterations $t \in 1, ..., T$}
            \State $\mathcal{A}_{t}$ = $\mathcal{M}(\mathbf{p},a^*_{t-1},f^*_{t-1},s^*_{t-1})$ \Comment{Refine algorithm (\autoref{reasoning_action})}
            \vspace{0.5em}
            
            \ForAll {objective vectors $\boldsymbol{\lambda} \in \{\Lambda^{n-1} \text{ discretized by } \Delta\lambda\}$} \Comment{Simplex (\autoref{simplex})}
                \State $Y_{\boldsymbol{\lambda}}=\mathcal{A}_t(x,\boldsymbol{\lambda})$
            \EndFor
            \vspace{0.5em}
            
            \State $Y_{t}$ $=$ \Call{Aggregate}{$Y_{\boldsymbol{\lambda} \in \Lambda}$}
            \vspace{0.5em}
            
            \If{\Call{Feasible}{$Y_t$}} \Comment{Feasibility (e.g., \autoref{budget}-\ref{binary})}
                \State $H_{t} \subseteq Y_{t}$ \Comment{Non-dominated set (\autoref{dominance}-\ref{H})}
                \State $\{f_{t},s_{t}\}=\text{IGD}(P,H_{t})$ \Comment{IGD metric (\autoref{IGD})}
            \Else
                \State $\{f_{t},s_{t}\}=\infty$
            \EndIf
            \vspace{0.5em}
            
            \If{$s_{t} \leq s^*_{t-1}$} \Comment{Current best iteration}
                \State $\{a^*_{t},f^*_{t},s^*_{t}\}$ $=$ \Call{Textify}{$\mathcal{A}_t,f_t,s_t$}
            \Else
                \State $\{a^*_{t},f^*_{t},s^*_{t}\}$ $=$ \Call{Textify}{$a^*_{t-1},f^*_{t-1},s^*_{t-1}$}
            \EndIf
        \EndFor
        \vspace{0.5em}
        
        \State\Return $a^*_T$
    
    \end{algorithmic}
    \end{algorithm}
    \caption{Agent framework algorithm.}
    \label{Moo--Agent}
\end{figure}

Similarly, regular reasoning-action \textsc{MoCo--Agent} iterations, from the first regular iteration $t=1$ to the final iteration $t=T$, employ the $p_{\text{PF}}$, $p_{\text{RA}}$, and $p_{\text{I/O}}$ engineered prompts, but incorporate information pertaining to best-scoring previous algorithm $a_t^*$, feedback $f_t^*$, and scoring $s_t^*$ using the prompt template $p$ for general iterating instructions. Here, $a_t^*$ is the \textit{"textified"} version of the best-scoring Python algorithm. Furthermore, its corresponding feedback $f_t^*$ contains the feasibility and number of non-dominated CCPO solutions, as well as any execution errors. The IGD score $s_t^*$ is equally included. The resulting engineered prompts are inputs to the language model $\mathcal{M}$ to generate a new algorithm $\mathcal{A}_t$ (\autoref{reasoning_action}).

Approximate solutions to the CCPO can then be generated using the Python metaheuristic algorithm $\mathcal{A}_t$ across the weight simplex (\autoref{simplex}) using problem-specific inputs $x$, e.g., the cardinality constraint parameter $K$ and the asset universe in the CCPO. For simplicity, the unit weight simplex $\Lambda^{n-1}$ is discretized uniformly by the step size $\Delta\lambda$ across all $n$-objective directions. Hence, in $n=2$ dimensions, the resulting weight vector $\boldsymbol{\lambda}=\{\lambda,1-\lambda\}$ is swept between $\lambda\in[0,1]$. Approximate solutions of the CCPO are aggregated across all sampled weight vectors $\boldsymbol{\lambda}$ into a solution set $Y_t$. The subset $H_t \subseteq Y_t$ of non-dominated solutions can be extracted (\autoref{dominance}-\ref{H}), after checking solution feasibility with respect to CCPO constraints (\autoref{budget}-\ref{binary}). Finally, the efficient frontier performance score $s_t$ can be measured with the IGD metric (\autoref{IGD}) using the standard Markowitz portfolio optimization UEF as a theoretically optimal reference. Here, $a_t^*$, $f_t^*$, and  $s_t^*$ are updated if the current iteration's algorithm $\mathcal{A}_t$ score $s_t$ is found to outperform the historical best $s_t^*$. In the general case where reference frontiers are not easily attainable, the hypervolume indicator \citep{HV} can be used as a suitable alternative metric. 

\subsection{Study Design}
The OpenAI o4-mini LLM is cast as a coding agent in this study using the \textsc{MoCo--Agent} framework. This study employs the \textsc{MoCo--Agent} in the generation of an algorithm portfolio for the CCPO. Here, multiple algorithms are generated by modifying the role assignment prompt $p_{\text{RA}}$ across this study's multiple instances of the \textsc{MoCo--Agent} algorithm. However, given the innumerable quantity of hitherto existing metaheuristic algorithms \citep{meta}, as well as the criticisms that recent metaphor-based metaheuristics have faced \citep{meta_crit}, this study limits itself to the generation of a small number of algorithms based on classical and early metaheuristics methods.

This study explores the generation of 10 algorithms, each based on a unique metaheuristic method. Evolutionary-based methods such as Genetic Algorithms (GA) \citep{GA}, Population-Based Incremental Learning (PBIL) \citep{PBIL}, and Differential Evolution (DE) \citep{DE} are included. Early nature-inspired methods such as Artificial Bee Colony (ABC) \citep{ABC} and Firefly Algorithm (FA) \citep{FA} are included. Swarm-based algorithms such as Particle Swarm Optimization (PSO) \citep{PSO} are included. Finally, trajectory-based methods such as Hill Climbing (HC) \citep{HC}, Simulated Annealing (SA) \citep{SA}, Tabu Search (TS) \citep{TS}, and Greedy Randomized Adaptive Search Procedure (GRASP) \citep{GRASP} are included. Furthermore, beyond specifying the metaheuristic method to be adapted by the \textsc{MoCo--Agent} in its reasoning step, the role assignment prompt $p_{\text{RA}}$ equally indicates that the coding agent should aim to hybridize the generated algorithm however it sees fit, but should not employ exact solution methods.

In the action step of the \textsc{MoCo--Agent}, the performance of the generated algorithms is measured on an asset universe taken from OR-Library \citep{OR-lib}. This study utilizes the same Hang Seng, DAX, FTSE, S\&P, and Nikkei asset universes used in foundational works for the CCPO \citep{CCPO}. Hence, for ease of comparison with the swath of following literature on CCPO, the same $K=10$, $\varepsilon=0.01$, and $\delta=1$ constraint parameters are taken, as well as $\Delta\lambda=0.02$ objective step size. The Hang Seng asset universe data, composed of 31 assets, is the smallest of the series taken from OR-Library, and is consequently selected as the training set used in algorithm generation. Each generated algorithm is given $T=32$ reasoning-action iterations, where an execution time limit of 10 minutes is given at each action iteration for each iteration's generated Python algorithm. The remaining DAX, FTSE, S\&P, and Nikkei asset universes are reserved as testing sets for the best-scoring algorithms $a_T^*$ generated using the Hang Seng training set.

\section{Results}
\subsection{Generated Algorithms}
The performance summary of the 10 metaheuristic algorithms generated with the \textsc{MoCo--Agent} framework is presented in \autoref{tab:gen_algs}, when measured on the Hang Seng training set. The generated algorithms are scored on two performance metrics. The first metric, being the Mean Percentage-deviation Error (MPE) \citep{CCPO}, is an indicator of efficient frontier convergence with respect to the UEF, and is measured on the solution set $V(\lambda)$. This set represents the solutions associated with the best objective function value for each sampled weight vector $\boldsymbol{\lambda}$ discretized by $\Delta\lambda$ steps. Hence, the set $V(\lambda)$ is comprised of 51 investment portfolio solutions as $\lambda\in[0,1]$ is sampled at every $\Delta\lambda=0.02$ step. Furthermore, dominated solutions are not eliminated. The second performance metric is the IGD \citep{IGD}, which is an indicator of both efficient frontier convergence and coverage.

\begin{table}[h]
    \centering
    \caption{Metaheuristic algorithm performance on the Hang Seng $(N=31)$ training set.}
    \resizebox{\textwidth}{!}{%
    \begin{tabular}{rcccccccccc}
        \toprule
         & \textbf{ABC} & \textbf{DE} & \textbf{FA} & \textbf{GA} & \textbf{GRASP} & \textbf{HC} & \textbf{PSO} & \textbf{PBIL} & \textbf{SA} & \textbf{TS} \\
        \midrule
        MPE & 1.0983 & 1.0965 & 3.4104 & 1.9444 & 1.2319 & 1.0965 & 1.1165 & 4.5653 & 1.4103 & 20.4934 \\
        \vspace{0.5em}
        Rank & 3 & 2 & 8 & 7 & 5 & 1 & 4 & 9 & 6 & 10 \\

        IGD & 0.53E-4 & 0.69E-4 & 1.61E-4 & 1.17E-4 & 0.88E-4 & 1.09E-4 & 0.70E-4 & 3.31E-4 & 1.43E-4 & 18.10E-4\\
        \vspace{0.5em}
        Rank & 1 & 2 & 8 & 6 & 4 & 5 & 3 & 9 & 7 & 10\\

        Score & 2 & 2 & 8 & 6.5 & 4.5 & 3 & 3.5 & 9 & 6.5 & 10\\
        \midrule
        Retain & Yes & Yes & No & No & Yes & Yes & Yes & No & No & No\\
        \bottomrule
    \end{tabular}}
    \label{tab:gen_algs}
\end{table}

For both performance indicators, the generated algorithms based on the FA, PBIL, and TS methods performed significantly worse than the others, while GA and SA also lagged behind slightly when measured on the Hang Seng training set. Hence, the ABC, DE, GRASP, HC, and PSO are retained for evaluation against the testing set, while the remainder are pruned. The generated algorithm portfolio is equally comprised of the four present metaheuristic method categories: evolutionary-based, nature-inspired, swarm-based, and trajectory-based. Furthermore, self-assessed descriptions of the generated algorithms are shown in \autoref{tab:descriptions}, a byproduct of the engineered prompts, which require the language model produce a self-assessment of the generated model at each reasoning iteration.

\begin{table}[h]
    \centering
    \caption{Self-assessed descriptions of generated algorithms.}
    \begin{tabular}{lp{10cm}r}
        \toprule
        Algorithm \hspace{1em} & Description \\
        \midrule
        \vspace{0.35em}
        \textbf{ABC} & \textit{"Enhanced ABC with biased initialization, adaptive neighbor strategies, and asset-frequency guidance".} \\
        \vspace{0.35em}
        \textbf{DE} & \textit{"Hybrid adaptive differential evolution with memetic local search including gradient-based weight adjustment and greedy asset swap moves".} \\
        \vspace{0.35em}
        \textbf{GRASP} & \textit{"Enhanced GRASP with cluster-diverse RCL, adaptive large neighborhood search operators, and operator selection learning".} \\
        \vspace{0.35em}
        \textbf{HC} & \textit{"Hybrid memetic hill-climbing with adaptive multi-swap moves, greedy swaps, crossover diversification, and tabu tenure".} \\
        \textbf{PSO} & \textit{"Hybrid PSO with ring topology, Levy-flight mutation, enhanced DE and adaptive local swap search".} \\
        \bottomrule
    \end{tabular}
    \label{tab:descriptions}
\end{table}

\subsection{Algorithm Portfolio Performance}
The generated algorithm portfolio is evaluated on the DAX, FTSE, S\&P, and Nikkei asset universes, which compose this study's testing sets. These datasets, taken from OR-Library \citep{OR-lib}, serve adequately as test sets as they differ substantially from the Hang Seng training set, both in terms of asset content and universe size. For example, the Hang Seng universe has $N=31$ assets, whereas the Nikkei universe has $N=225$ assets. The performance against the testing set is compared equally against the State Of The Art (SOTA). \autoref{tab:comp_SOTA} presents the performance of the generated metaheuristic algorithms on the training and testing sets. Furthermore, the performance of the SOTA is included as a reference. Performance is measured using the Percentage-deviation Error (PE) indicator \citep{CCPO} using the asset universes' respective UEFs as reference theoretical optimal frontiers, where mean, median, min, and max PE for sets $V(\lambda)$ are reported across all included algorithms.

\begin{table}[h]
    \centering
    \caption{Comparison with SOTA for $V(\lambda)$.}
    \resizebox{\textwidth}{!}{%
    \begin{tabular}{llcccccc}
        \toprule
        \hspace{10em} & \hspace{4em} &\textbf{ABC} & \textbf{DE} & \textbf{GRASP} & \textbf{HC} & \textbf{PSO} & \textbf{SOTA}$^{\dagger}$\\
        \midrule
        Hang Seng $(N=31)$ & Mean & $1.0983$ & $\mathbf{1.0965}$ & $1.2319$ & $\mathbf{1.0965}$ & $1.1165$ & $\underline{1.0873}$\\
        & Median & $\mathbf{\underline{1.2149}}$ & $1.2155$ & $1.2155$ & $1.2154$ & $1.2211$ & $1.2154$\\
        & Min & $\mathbf{\underline{0.0000}}$ & $\mathbf{\underline{0.0000}}$ & $\mathbf{\underline{0.0000}}$ & $\mathbf{\underline{0.0000}}$ & $0.0224$ & $\underline{0.0000}$\\
        \vspace{0.5em}
        & Max & $1.5538$ & $1.5538$ & $8.4637$ & $\mathbf{\underline{1.5537}}$ & $1.6241$ & $1.5538$\\
        
        DAX  $(N=85)$ & Mean & $2.5920$ & $\mathbf{2.3232}$ & $2.3778$ & $2.3398$ & $2.5802$ & $\underline{2.2898}$\\
        & Median & $2.6050$ & $\mathbf{\underline{2.5470}}$ & $2.5630$ & $2.5630$ & $2.6069$ & $2.5629$\\
        & Min & $\mathbf{\underline{0.0059}}$ & $\mathbf{\underline{0.0059}}$ & $\mathbf{\underline{0.0059}}$ & $\mathbf{\underline{0.0059}}$ & $0.0164$ & $\underline{0.0059}$\\
        \vspace{0.5em}
        & Max & $10.3078$ & $5.7669$ & $6.7775$ & $\mathbf{4.3210}$ & $5.8620$ & $\underline{4.0275}$\\

        FTSE $(N=89)$ & Mean & $1.0042$ & $0.9443$ & $0.9728$ & $\mathbf{0.8799}$ & $1.1566$ & $\underline{0.8406}$\\
        & Median & $\mathbf{\underline{1.0841}}$ & $\mathbf{\underline{1.0841}}$ & $\mathbf{\underline{1.0841}}$ & $\mathbf{\underline{1.0841}}$ & $1.2485$  & $\underline{1.0841}$\\
        & Min & $0.0045$ & $\mathbf{\underline{0.0016}}$ & $\mathbf{\underline{0.0016}}$ & $0.0327$ & $0.0327$ & $\underline{0.0016}$\\
        \vspace{0.5em}
        & Max & $4.9484$ & $5.9721$ & $7.5422$ & $\mathbf{2.6819}$ & $2.8589$ & $\underline{2.0670}$\\
        
        S\&P $(N=98)$ & Mean & $1.7416$ & $1.8478$ & -- & $\mathbf{1.4351}$ & $1.6552$ & $\underline{1.3464}$\\
        & Median & $1.1810$ & $1.2144$ & -- & $\mathbf{\underline{1.1420}}$ & $1.1755$ & $1.1515$\\
        & Min & $0.0046$ & $0.2655$ & -- & $\mathbf{\underline{0.0009}}$ & $0.0061$ & $\underline{0.0009}$\\
        \vspace{0.5em}
        & Max & $9.8889$ & $8.9263$ & -- & $\mathbf{7.7908}$ & $10.5647$ & $\underline{5.4520}$\\

        Nikkei $(N=225)$ & Mean & $0.6787$ & $2.7516$ & $0.7184$ & $\mathbf{0.5782}$ & $1.2403$ & $\underline{0.5665}$\\
        & Median & $0.6229$ & $1.8011$ & $0.6118$ & $\mathbf{\underline{0.5858}}$ & $1.1422$ & $\underline{0.5858}$\\
        & Min & $\mathbf{\underline{0.0000}}$ & $0.5863$ & $0.0238$ & $0.0003$ & $0.1675$ & $\underline{0.0000}$\\
        & Max & $2.7825$ & $7.9270$ & $5.0087$ & $\mathbf{\underline{1.1606}}$ & $3.1126$ & $\underline{1.1606}$\\
        \bottomrule
    \end{tabular}}
    \begin{flushleft}
    \footnotesize
    $^{\dagger}$ ABC-HP algorithm \citep{ABC-HP} measured as the best performing in the SOTA \citep{SOTA}.
    \end{flushleft}
    \label{tab:comp_SOTA}
\end{table}

The score(s) of the best-performing algorithm(s) are \underline{underlined} in \autoref{tab:comp_SOTA}, whereas the score(s) of the best-performing algorithm(s) generated via the \textsc{MoCo--Agent} framework are \textbf{emboldened}. Metrics for GRASP are left empty for the S\&P test set, as the Python algorithm code execution time limit of 30 minutes was reached before portfolios were found across all sampled $\Delta\lambda$ steps. \autoref{tab:comp_SOTA} indicates that the generated algorithm portfolio frequently matches SOTA performance across several performance indicators, and can also infrequently surpass SOTA performance. Here, the SOTA is taken as the High-Population ABC (ABC-HP) \citep{ABC-HP} algorithm as measured in \citep{SOTA}. However, the generated algorithm portfolio greatly outperforms historical SOTA algorithms measured on $V(\lambda)$ using PE indicators such as the GA, TS, and SA from \citep{CCPO}, PSO from \citep{PSO_hist}, and PBIL-DE from \citep{PBIL_DE}.

\subsection{Pooled Metaheuristics}
\begin{figure}[h]
    \centering
    
    \begin{minipage}{0.49\textwidth}
        \centering
        \includegraphics[width=\linewidth]{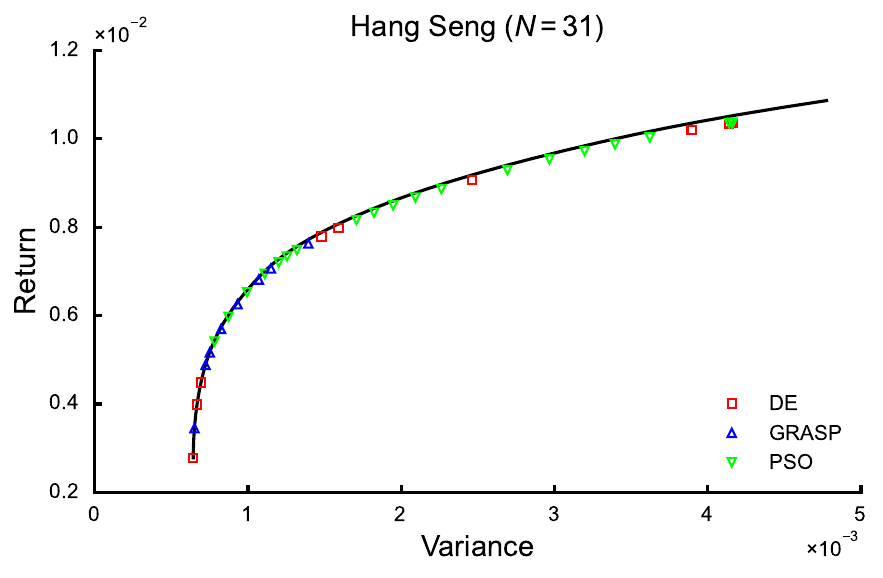}
    \end{minipage}\hfill
    \begin{minipage}{0.49\textwidth}
        \centering
        \includegraphics[width=\linewidth]{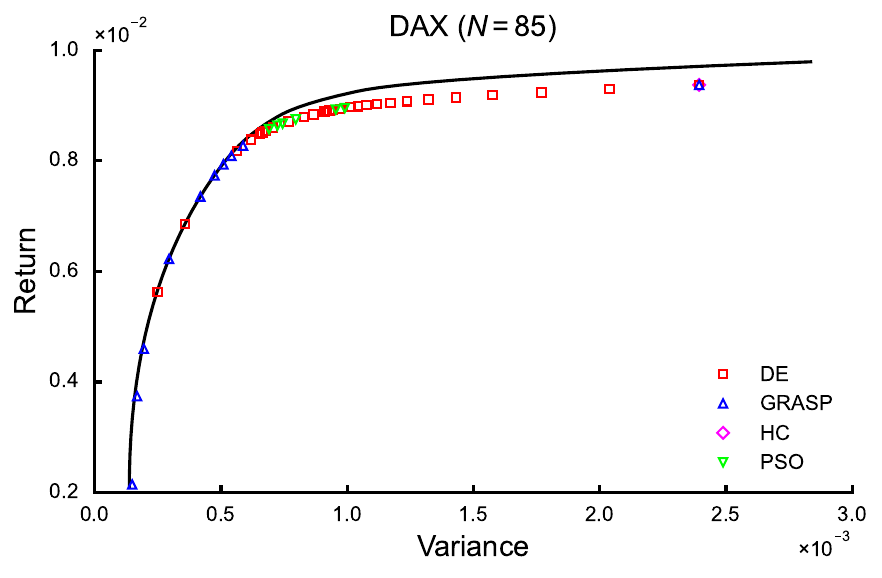}
    \end{minipage}

    \vspace{1em}

    \begin{minipage}{0.49\textwidth}
        \centering
        \includegraphics[width=\linewidth]{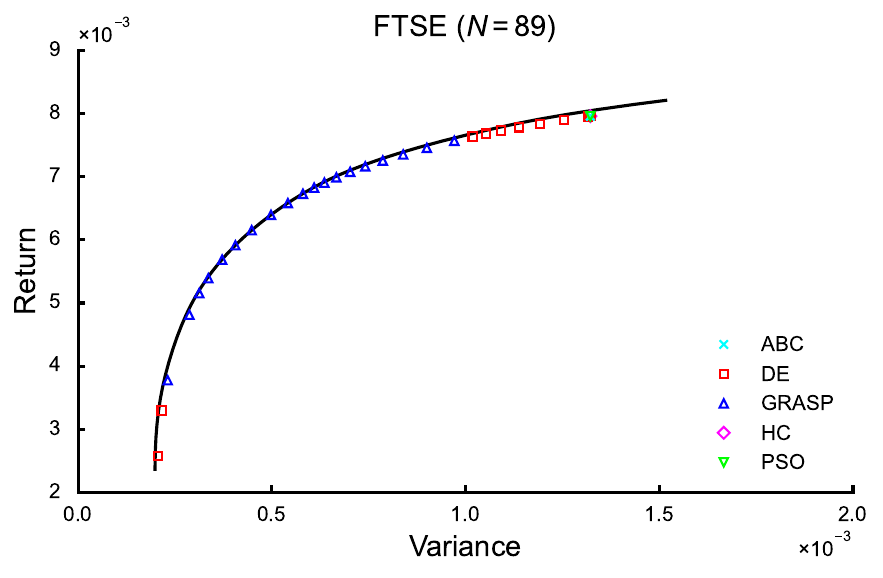}
    \end{minipage}\hfill
    \begin{minipage}{0.49\textwidth}
        \centering
        \includegraphics[width=\linewidth]{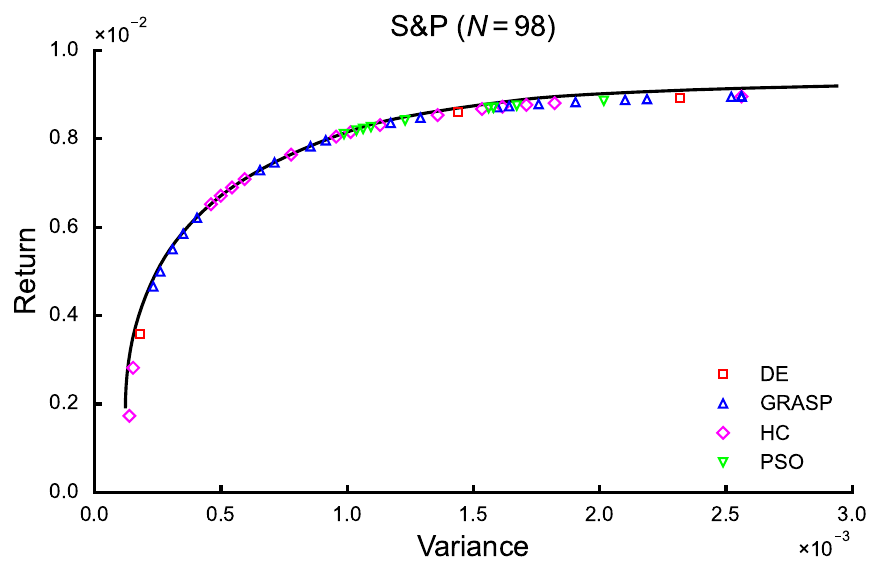}
    \end{minipage}

    \vspace{1em}

    \begin{minipage}{0.49\textwidth}
        \centering
        \includegraphics[width=\linewidth]{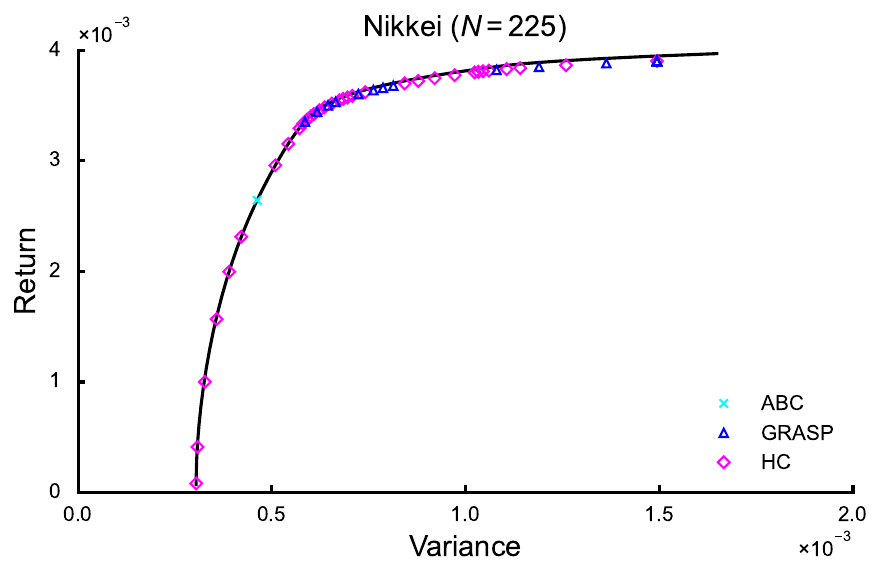}
    \end{minipage}

    \caption{Pooled metaheuristic $V(\lambda)$ frontiers.}
    \label{fig:benchmarks}
\end{figure}

Beyond individual algorithm performance, the solutions of the generated algorithm portfolio can be pooled to improve efficient frontier performance. \autoref{fig:benchmarks} illustrates the efficient $V(\lambda)$ frontier of the pooled investment portfolio solutions from the generated metaheuristic algorithms across the Hang Seng, DAX, FTSE, S\&P, and Nikkei asset universes taken from OR-Library \citep{OR-lib}. The black lines in \autoref{fig:benchmarks} represent the datasets' respective UEFs. The best algorithm portfolio solution is taken for each sampled $\lambda$ step. However, dominated solutions across different sampled $\lambda$ values are not eliminated. \autoref{fig:benchmarks}, along with \autoref{tab:cont_front}, indicates the variety across the extent to which generated metaheuristic algorithms contribute to the efficient frontiers, where algorithms differ significantly in terms of their respective contributions across asset universes. For example, the DE and PSO algorithms contribute greatly to the Hang Seng frontier, whereas the HC algorithm makes no contribution. Conversely, the HC algorithm contributes greatly to the Nikkei frontier, whereas the DE and PSO algorithms make no contribution. Notably, the GRASP algorithm makes important contributions to the frontiers of each evaluated asset universe.

\begin{table}[h]
    \centering
    \caption{Metaheuristic algorithm contributions to pooled $V(\lambda)$ frontiers.}
    \begin{tabular}{lccccc}
        \toprule
        \hspace{10em} & \textbf{ABC} & \textbf{DE} & \textbf{GRASP} & \textbf{HC} & \textbf{PSO} \\
        \midrule
        \vspace{0.35em}
        Hang Seng $(N=31)$ & -- & $19/51$ & $8/51$ & -- & $24/51$ \\
        \vspace{0.35em}
        DAX  $(N=85)$ & -- & $34/51$ & $10/51$ & $1/51$ & $6/51$ \\
        \vspace{0.35em}
        FTSE $(N=89)$ & $3/51$ & $27/51$ & $19/51$ & $1/51$ & $1/51$ \\
        \vspace{0.35em}
        S\&P $(N=98)$ & -- & $3/51$ & $23/51$ & $16/51$ & $9/51$ \\
        Nikkei $(N=225)$ & $1/51$ & -- & $17/51$ & $33/51$ & -- \\
        \bottomrule
    \end{tabular}
    \label{tab:cont_front}
\end{table}

The contribution of individual algorithms within the generated algorithm portfolio can also be measured by their contributions towards improving the IGD indicator. \autoref{tab:cont_IGD} indicates the relative improvement of the pooled IGD when specific algorithm solutions are incorporated. For example, the pooled IGD indicator is improved by $48.52\%$ on the Hang Seng asset universe when the ABC solutions are added to the pooled DE, GRASP, HC, and PSO solutions. \autoref{tab:cont_IGD} also indicates the variety across the extent to which algorithms contribute to the improvement of the IGD indicator. Unlike the PE metrics, the IGD indicator not only measures convergence but also measures coverage. For example, the ABC algorithm, which made almost no contribution to the evaluated $V(\lambda)$ frontiers, contributed significantly to improving IGD across all evaluated asset universes. This indicates that while its per $\lambda$ convergence performance might be relatively poor, the coverage of its set $H$ of non-dominated solutions is very good. Conversely, algorithms such as DE or GRASP, which made important contributions to the $V(\lambda)$ frontiers, had little impact on the IGD metric across several of the evaluated asset universes. This illustrates the importance of the diversity of algorithms contained within the generated algorithm portfolio, where different algorithms respectively have varying levels of contribution to efficient frontier convergence and coverage.

\begin{table}[h]
    \centering
    \caption{Metaheuristic algorithm contributions to improving pooled IGD.}
    \begin{tabular}{lccccc}
        \toprule
        \hspace{10em} & \textbf{ABC} & \textbf{DE} & \textbf{GRASP} & \textbf{HC} & \textbf{PSO} \\
        \midrule
        \vspace{0.35em}
        Hang Seng $(N=31)$ & $48.52\%$ & $0.12\%$ & $0.03\%$ & $4.42\%$ & $0.04\%$ \\
        \vspace{0.35em}
        DAX  $(N=85)$ & $33.21\%$ & $10.38\%$ & $16.80\%$ & $2.09\%$ & $35.05\%$ \\
        \vspace{0.35em}
        FTSE $(N=89)$ & $30.32\%$ & $11.08\%$ & $34.53\%$ & $2.16\%$ & $3.32\%$ \\
        \vspace{0.35em}
        S\&P $(N=98)$ & $61.04\%$ & $8.89\%$ & $0.58\%$ & $118.28\%$ & $7.62\%$ \\
        Nikkei $(N=225)$ & $215.08\%$ & $0\%$ & $5.63\%$ & $32.54\%$ & $18.22\%$ \\
        \bottomrule
    \end{tabular}
    \label{tab:cont_IGD}
\end{table}

%\section{Brief Discussion}
\section{Conclusion}
In conclusion, this study successfully implements an agentic framework for the generation of algorithm portfolios for multi-objective combinatorial optimization. The framework is applied to the NP-hard CCPO problem, where a portfolio of metaheuristic algorithms is successfully generated without extensive human developmental effort. The resulting algorithm portfolio is validated across benchmark asset universes and compared with the SOTA. The findings show that algorithms generated from the agentic framework frequently match, and in some cases surpass, the performance of the SOTA. Furthermore, the study also investigates the impact of pooling solutions from the generated algorithms, revealing that each algorithm contributes uniquely to different regions of efficient frontier convergence and coverage. The study is limited to the CCPO problem and to one agent framework based on greedy refinement. Future research should explore alternative agentic framework designs and extend the evaluation to other multi-objective combinatorial optimization problems to further validate performance. Furthermore, future research should consider the interaction between different generated algorithms, where the efficiency of efficient frontier generation can be improved via hyper-heuristics. Nevertheless, the results demonstrate the effectiveness of the proposed agentic framework in generating algorithms for the CCPO and demonstrate the overall developmental utility of the implemented agentic framework, where the diversity of derived algorithms in pooling can improve the global performance of the studied efficient frontiers.

\section*{Acknowledgments}
This work was supported by the Natural Sciences and Engineering Research Council of Canada (NSERC) Discovery Grant, and the McGill Engineering Doctoral Award (MEDA). Compute resources were provided by the Digital Research Alliance of Canada and Calcul Qu\'ebec.

\bibliographystyle{unsrtnat}
\bibliography{ref}

\end{document}